\documentstyle[12pt]{article}
\begin{document}
\pagestyle{empty}
\noindent
MPG-VT-UR 210/00
\hfill{\sf Hot Points in Astrophysics}

\hfill{\sf JINR, Dubna, Russia, August 22-26, 2000}

\vspace*{2.cm}

\centerline{\bf Cosmological Consequences of Conformal General Relativity}

\vspace*{0.2cm}

\begin{center}
Danilo Behnke$^{\dagger}$, David Blaschke$^{\dagger}$,  \\
Victor Pervushin$^{\ddagger}$, Denis Proskurin$^{\ddagger}$ and
Alexandr Zakharov$^*$\\[0.3cm]

{\it $^\dagger$Rostock University, 18051 Rostock, Germany\\
$^{\ddagger}$Joint Institute for Nuclear Research, 141980 Dubna, Russia\\
$^{*}$Institute of Theoretical and Experimental Physics, 117259
Moscow, Russia}

\end{center}

\vspace*{0.2cm}

\begin{abstract}
We consider cosmological consequences of a conformal-invariant
unified theory which is dynamically equivalent to general relativity
and is given in a space with the geometry of similarity. We show that
the conformal-invariant theory offers new explanations for 
to such problems as the arrow of time, initial cosmic data, dark matter and
accelerating evolution of the universe in the dust stage.
\end{abstract}

\vspace*{0.5cm}

\def\btt#1{{\tt$\backslash$#1}}
\def\v1{\vspace{1cm}}
\def\be{\begin{equation}}
\def\ee{\end{equation}}
\def\bc{\begin{center}}
\def\ec{\end{center}}
\def\ik{\partial}
\def\vh{\varphi}
\def\ve{\varepsilon}
\newcommand{\ap}{\approx}
\newcommand{\ch}{\check}
\newcommand{\vth}{\vartheta}
\newcommand{\ih}{\imath}
\newcommand{\sig}{\Sigma}
\newcommand{\p}{\underline}
\newcommand{\fr}{\frac}
\newcommand{\vph}{\varphi}
\newcommand{\lm}{\lambda}
\newcommand{\pr}{\partial}
\newcommand{\kp}{\kappa}
\newcommand{\sq}{\sqrt}
\newcommand{\de}{\delta}
\newcommand{\al}{\alpha}
\newcommand{\lf}{\left}
\newcommand{\ri}{\right}
\newcommand{\ol}{\overline}
\newcommand{\im}{\imath}
\newcommand{\cH}{{\cal H}}
\newcommand{\bea}{\begin{eqnarray}}
\newcommand{\eea}{\end{eqnarray}}
\renewcommand{\thesection}{\arabic{section}.}
\renewcommand{\thesubsection}{\thesection\arabic{subsection}.}

\section*{Introduction}

There are observations~\cite{pr}-\cite{kl} that the classical equations of
Einstein's general relativity (GR) are dynamically equivalent to
the conformal-invariant theory described by the
Penrose-Chernikov-Tagirov-type action~\cite{pct} with a negative sign
for the scalar (dilaton) field $\Phi$ referred to as a conformal compensator. 
This dilaton version of GR (considered also as a particular case of the 
Jordan-Brans-Dicke scalar tensor theory of gravitation~\cite{jbd})
is the basis of some speculations on the unification of Einstein's gravity  
with the Standard Model of electroweak and strong interactions~\cite{pr,plb,kl}
including modern  theories of supergravity~\cite{kl}.
In the conformal-invariant Lagrangian of matter, the dilaton generates the 
masses of elementary particles, i.e. it plays the role of the modulus of the 
Higgs field.

However, in the current literature ~\cite{kl} a peculiarity of the 
conformal-invariant version of Einstein's dynamics has been overlooked.
The conformal-invariant version of Einstein's dynamics
is not compatible with the absolute standard of
measurement of lengths and times given by an interval in the
Riemannian geometry as this interval is not conformal-invariant.
As it has been shown by Weyl in 1918~\cite{we}, conformal-invariant
theories correspond to the relative standard of measurement of
a conformal-invariant ratio of two intervals,
given in the geometry of similarity as a manifold of Riemannian geometries
connected by conformal transformations.
The geometry of similarity is characterized by a measure of changing
the length of a vector in its parallel transport. In the considered
dilaton case, it is the gradient of the dilaton $\Phi$~\cite{plb}.
In the following, we call the scalar conformal-invariant theory
the conformal general relativity (CGR) to distinguish it from the original
Weyl~\cite{we} theory where the measure of changing the length
of a vector in its parallel transport is a vector field
(that leads to the defect of the physical ambiguity of the arrow of time
pointed out by Einstein in his comment to Weyl's paper~\cite{we}).
Thus, the choice between two dynamically equivalent theories ---
general relativity (GR) and conformal general relativity
(CGR) is the choice between the Riemannian
geometry and Weyl's geometry of similarity.
Two different geometries for the same dynamics correspond to different
standards of measurement and two different cosmological pictures
for different observers:
(I) an Einstein observer, who supposes that he measures an absolute interval
of the Riemannian geometry,
obtains the Friedmann-Robertson-Walker (FRW) cosmology where the
redshift is treated as expansion of the universe;
(II) a Weyl observer, who supposes that he measures a relative interval
of the geometry of similarity,
obtains a field version of the Hoyle-Narlikar cosmology \cite{N}.
The redshift and the Hubble law in  the Hoyle-Narlikar cosmology \cite{N}
reflect the change of the size of atoms in the process of evolution
of masses of elementary particles generated by the scalar dilaton field
\cite{grg,plb,N}.

The present paper is devoted to a discussion of cosmological consequences
of the conformal-invariant dilaton gravity.

\section*{Conformal General Relativity}

%\subsection*{Action}

We start from the conformal-invariant theory described
by the sum of the dilaton action and the matter action
\be \label{cut}
W=W_{\rm CGR}+W_{\rm matter}.
\ee
The dilaton action is the Penrose-Chernikov-Tagirov one for a scalar
(dilaton) field with the negative sign
\be
\label{wgr}
W_{\rm CGR}(g|\Phi)= \int d^4x\left[-\sqrt{-g}\frac{\Phi^2}{6} R(g)+
\Phi \partial_{\mu}(\sqrt{-g}g^{\mu\nu}\partial_{\nu}\Phi )\right]~.
\ee
The conformal-invariant action of the matter fields can be chosen in the form
\be
\label{smc}
W_{\rm matter}=\int d^4x\left[{\cal L}_{(\Phi=0)}
+\sqrt{-g}(-\Phi F+\Phi^2B-\lambda \Phi^4)\right]~,
\ee
where $B$ and $F$ are the mass contributions to the Lagrangians of the vector 
($v$) and spinor ($\psi$) fields, respectively,
\be \label{67}
B=v_i (y_v)_{ij}v_j~;~~
F=\bar\psi_{\alpha} (y_s)_{\alpha\beta}\psi_{\beta},
\ee
with $(y_v)_{ij}$, and $(y_s)_{\alpha\beta}$ being the mass matrices of 
vector bosons and fermions coupled to the dilaton field.
The massless part of the Lagrangian density of
the considered vector and spinor fields is denoted by ${\cal L}_{(\Phi=0)}$.
The class of theories of the type ~(\ref{cut}) includes the
superconformal theories with supergravity~\cite{kl} and the standard model 
with a massless Higgs field \cite{plb} as the mass term would violate the 
conformal symmetry.

%\subsection*{Conformal - invariant variables}

Following Dirac we suppose that the symmetry of the theory
establishes the symmetry of its observable quantities. In other words,
the conformal invariance of the theory entails the
conformal invariance of physical variables and measurable quantities.

Recall that in GR the problems of initial data and time evolution are
studied with the help of the Lichnerowicz conformal-invariant variables
$F^L$~\cite{L}
\be \label{lich}
{}^{(n)}F^L={}^{(n)}F|{}^{3}g|^{-n/6},
\ee
where ${}^{(n)}F$ is a set of fields with a conformal weight $(n)$
including the metric $g_{\mu\nu}$ with $|{}^{3}g^L_{ij}|=1 $.
The Lichnerowicz variables~(\ref{lich}) are defined by
the Dirac-ADM-foliation of the metric
\be
\label{dse}
  (ds^L)^2=g^L_{\mu\nu}dx^\mu dx^\nu=(N^{L} dt)^2-{}^{(3)}g^L_{ij}
\breve{dx}{}^i \breve{dx}{}^j\;,~~~~~\;\;\breve{dx}{}^i=dx^i+N^idt
\ee
with the lapse function $N^L(t,\vec x)$, three shift vectors $N^i(t,\vec x)$,
and five space components ${}^{(3)}g^L_{ij}(t,\vec x)$ depending on the
coordinate time $t$ and the space coordinates $\vec x$.
This definition excludes one superfluous degree of
freedom and gives a conformal - invariant "measurable" interval in
the CGR theory considered.

In terms of the Lichnerowicz variables~(\ref{lich}),
the dynamic equivalence of GR and CRG becomes evident
\be \label{equi}
W_{\rm CGR}[g^L|\Phi^L]=W_{\rm GR}[g^L|\Phi_{\rm scale}]~.
\ee
Einstein's theory is obtained by replacing the
Lichnerowicz dilaton field $\phi^L$ with the new scale field
\be \label{crgr}
\Phi_{\rm scale}\equiv M_{\rm Planck}\sqrt{\frac{3}{8\pi}}
|{}^{3}g|^{1/6} .
\ee
It can be possible by introducing the dimensional constant in
the conformal -invariant theory~(\ref{cut}). Therefore,
the renunciation from the dimensional constant means the renunciation
from the Einstein definition of measurable intervals in GR.
Instead of the Einstein intervals we shall use the conformal -
invariant Lichnerowicz intervals~(\ref{dse}) without the determinant
of the spatial metric (that disappears in the ratio of two intervals).

The opinion dominates that a dimensional constant (of the type of the
Planck mass) can appear due to spontaneous conformal symmetry breaking
in quantum perturbation theory; it can be a reason for introducing
this constant  in the theory from the very beginning~\cite{pr,kl}.

The formulation of the consistent reparametrization - invariant
perturbation theory for Einstein's general relativity (GR)
in Refs.~\cite{grg,plb,ps1,6116} gave a set of arguments in
favor of the opposite point of view:
the reparametrization - invariant
perturbation theory does not violate the conformal symmetry. Therefore,
in CGR, the role of the Planck mass  is played by the dilaton field.

In the present paper we formulate conformal cosmology
as a particular case of the conformal - invariant and reparametrization -
invariant perturbation theory.

\section*{Conformal - Invariant Theory of Cosmic Evolution}

%\subsection*{Conformal - invariant perturbation theory}

In the considered case of CGR~(\ref{cut}) in terms of the conformal -
invariant fields~(\ref{lich}), perturbation theory begins from
the homogeneous approximation for the dilaton and the  metric
\be \label{crin0}
 \Phi^L(t,x)=\vh(t),~N^L(t,x)=N_0(t),~~~~~ g^L_{ij}=\delta_{ij} ~,
\ee
which conserves the reparametrization - invariance even in the case
of free conformal fields described by the action
\be \label{gradl}
W_0=\int\limits_{t_1 }^{t_2 }dt \left[
\vh \frac{d}{dt}\frac{\dot \vh}{N_0}V_0 + N_{0}L_{0}\right],
\ee
where $V_0$ is a finite spatial volume.
$L_0$ is the sum of the Lagrangians of free fields,
\be \label{H0l}
L_0=L_M+L_R+L_h,
\ee
where in particular
\be \label{hml}
L_M= \frac{1}{2} \int\limits_{V_0 } d^3x\left(\frac{{\dot v}^2}{N_0^2}-
(\ik_iv)^2 -
(y_v\vh)^2 v^2\right)+ \int\limits_{V_0 }
d^3x \left(\bar \psi(-y_s\vh-\frac{\gamma_0}{N_0}\ik_0
+i\gamma_j\ik_j)\psi \right)
\ee
is the Lagrangian of massive conformal fields~(\ref{lich})
(massive vector ($v$), and spinor ($s$)~).
The role of the masses is played by the homogeneous dilaton field
$\vh $ multiplied by dimensionless
constants $y_f$, where $f$ labels the particle species. 
$L_R$ is the Lagrangian of massless fields (photons $\gamma$, neutrinos $\nu$) 
with $y_{\gamma}=y_{\nu} =0$, and
\be \label{hol}
L_h=\int\limits_{V_0 }^{ }
d^3x\frac{\vh^2}{24}\left(\dot h^2-
(\ik_ih)^2\right),
\ee
is the Lagrangian of the gravitons ($h_{ii}=0;~~\ik_jh_{ji}=0$) as weak 
transverse excitations of spatial metric
(the last two equations follow from the unit determinant
of the three-dimensional metric~(\ref{lich}) and from the momentum constraint).

This reparametrization - invariant action~(\ref{gradl}) describes the
well-known system of the free conformal fields
in a finite space-volume used for studying the problem of
creation of particles by the homogeneous excitation of the
metric~\cite{ps1,lp,z}.

We propose that $V_0$ coincides with the volume of the whole universe,
so that a nonlocalizable energy~\cite{rt} does not appear.

We call the system~(\ref{gradl}) with the invariant geometric time
and the stationary metric
\be \label{geot}
dT=N_0dt,~~~~~~~~~~~~~~~~~~~~~~~~~~~   ds^2_0=dT^2 - dx_i^2,
\ee
a conformal - invariant universe. Our task is to find the evolution of
all fields in the field world space $(\vh,f)$ with respect to the
geometric time $T$.

%\subsection*{ Geometric time evolution }

The variation of the action with respect to
the homogeneous lapse-function $N_0$ yields the energy constraint,
\be
\label{d56}
\frac{\delta W_{0}}{\delta N_0}=0~
\Rightarrow\,
\left(\frac{d\vh}{N_0dt}\right)^2=\frac{H(\vh)}{V_0}:=\rho(\vh)~,
\ee
where
\be \label{H}
H=\frac{\delta L_{0}}{\delta N_0}
\ee
is the Hamilton function of all field excitations with positive energy.
Its expectation value determines the measurable energy density of all particles
including gravitons~(\ref{hol}), see below.
The solution of equation (\ref{d56}),
\be
\label{70}
T_{\pm}({\vh_0})=\pm\int\limits_0^{\vh_0}d\vh
{\rho}^{-1/2}(\vh),
\ee
describes the evolution of the dilaton
(i.e. of all masses) with respect to the geometric time $T$.

%\subsection*{The arrow of the geometric time}

Equation~(\ref{d56}) as an energy constraint
restricts the spectrum of values of the dilaton momenta
to two solutions (with a positive sign and a negative one).
In the equivalent unconstrained system (which can be constructed
by substituting the solution of the Abelian energy constraint~(\ref{d56})
into the extended action~\cite{grg,ps1}) the dilaton momentum plays the
role of the Hamiltonian of the evolution in $\vh$.

Recall that in order to obtain a stable quantum theory, according to Dirac, 
one should propose that a universe with a positive energy ($+$)
propagates forward with respect to the dynamic evolution
parameter ($\vh_0 > \vh_1$); and with a
negative energy ($-$), backward ($\vh_0 < \vh_1$). In both cases
the geometric time~(\ref{70}) is always positive
\be \label{posi}
T_+(\vh_0>\vh_1)>0,~~~~~~~~~~~~
T_-(\vh_0<\vh_1)>0~.
\ee
The quantization of the dilaton field and
the elimination of negative eigenvalues of the Hamiltonian
(i.e. negative energies) by the Dirac treatment
of the branch with the negative Hamiltonian as annihilation
of  universes with a positive energy immediately leads to
the arrow of geometric time~\cite{6116}.

%\subsection*{ Particles and quasiparticles }

We introduce the particles as holomorphic field variables
\be
\label{grep}
   f(t,\vec x)=\sum\limits_{f,k}^{ }
\frac{C_f(\vh)\exp(ik_ix_i)}{V_0^{3/2}\sqrt{2\omega_f(\vh,k)}}
\left( a_f^+(-k,t)+ a_f^-(k,t)\right)~,
\ee
where $\omega_{f}(\vh,k)=\sqrt{k^2+y_f^2\vh^2}$ are the one-particle energies 
for the particle species $f=h,\gamma,\nu,v,s$ with the dimensionless mass 
parameters $y_f$ and the coefficients $C_f(\vh)$ are
\be
C_h(\vh) = \vh\sqrt{12},~~~~C_{\gamma}=C_\nu=C_v=C_s=1~.
\ee
These variables diagonalize the operator of the observational
density of matter in the well-known QFT form
\be \label{dide}
\rho(\vh)=\frac{H(\vh)}{V_0}=
\sum\limits_{f, k }^{ } \frac{\omega_{f}(\vh,k)}{V_0} N_{f}(\vh,k)~,
\ee
where
\be \label{part}
{\cal N}_{a_f}=<|\{a_f^+,a_f\}_{\pm}|>=<|\frac{1}{2}(a_f^+a_f\pm a_f a^+_f))|>
\ee
is the expectation value of the number of particles
(the upper sign corresponds to bosons, the lower one to fermions), 
$<|$ and $|>$ is
the physical states determined by the initial cosmic data.
In the following, we restrict ourselves to gravitons and massive vector 
particles.

This definition of particles~(\ref{grep}) excludes the vertices
of the dilaton - matter coupling which can restore the Higgs-type
potential in the perturbation theory by the Coleman-Weinberg
summing of the perturbation series.
If we exclude from the very beginning the $\lambda \phi^4$ term
from the initial action in order to remove
a tremendous vacuum density~\cite{ws},
this term could not be restored by the perturbation series.
Therefore, we suppose in the following that $\lambda = 0$.

Solutions of the equations of motion corresponding to the action of
the system~(\ref{gradl}) - (\ref{hol}) can be obtained by a Bogoliubov 
transformation
\bea
b_f^+ = \cosh{(r_f)}e^{-i\theta_f}a_f^+ -i\sinh{(r_f)}e^{i\theta_f}a_f~,\\
b_f   = \cosh{(r_f)}e^{i\theta_f}a_f +i\sinh{(r_f)}e^{-i\theta_f}a_f^+~,
\eea
where $b_f^+$ and $b_f$
are the creation and annihilation operators of Bogoliubov quasiparticles with 
$N_{f}=\{b_{f}^+, b_{f}\}$ being the operator of the (conserved)
numbers of quasiparticles~\cite{ps1}.

%\subsection*{Cosmic initial state and creation of matter from "nothing"}

We choose the initial state appropriate to the integrals of motion
to be the quasiparticle vacuum state defined by $b|0>=0$. 
We call this state the "nothing" in order to distinguish it from the vacuum of
observable particles.
In this case, a set of equations for the expectation values of
the numbers of particles (gravitons and mesons) are~\cite{ps1,6116,ppv}
\bea
N_f &=& {1\over {2}} \cosh{(2r_f)}~ =~ <0|\{a_f^+,a_f\}|0>~,\\
(\omega_f-\theta_f')\sqrt{4N_f^2-1} &=&
\Delta_f\cos{(2\theta_f)}2N_f~,\\
N_f' &=& -\Delta_f\sin{(2\theta_f)}\sqrt{4N_f^2-1}~~~~~~~~,
\eea
where the dash denotes the derivative $d/dT$ with respect to the geometric 
time (\ref{geot}) and $\Delta_h =
{\varphi'}/{\varphi},~ \Delta_v ={\omega'}/{(2\omega)}$.

The equations for the coefficients of the Bogoliubov
transformations can be solved explicitly in two limits: 
at the beginning of
the Universe and at the present-day stage~\cite{ps1}.

At the beginning of the Universe in the state of the Bogoliubov
(i.e. squeezed) vacuum of quasiparticles, we got the density of measurable
gravitons (particles) which corresponds to the well-known anisotropic
(Kasner) stage with the Misner wave function of the Universe~\cite{M}.
The anisotropic stage is followed by the stage of inflation-like
increase in the cosmic scale factor with respect to the time measured
by an observer with the relative standard.
At these stages,
the Bogoliubov quasiparticles strongly differ from
the measurable particles.

\section*{Observational consequences of Conformal Cosmology }

%\subsection*{Evolution of mass in the Hoyle-Narlikar conformal cosmology}

The initial action~(\ref{gradl}) shows that all masses of particles
increase with the geometric time.
A photon emitted by an
atom of a star two billion years ago remembers the size of that atom
(i.e., its mass) at the time of emission ($T_0-D$). After two billion
years, at the time of detection on the earth ($T_0$), the wavelength
of a photon is compared to the wavelength of a photon emitted by a
standard atom on the Earth when its size decreased due to the cosmic
evolution of the masses of elementary particles. This is just
the version of cosmology proposed by Hoyle and Narlikar~\cite{N} where the 
origin of the redshift
\be \label{z}
Z(D)=\frac{\vh(T_0)}{\vh(T_0-D/c)}-1= {\cal H}_{0}(T_0) D/c+...~,
\ee
is the evolution of particle masses determined in our case by the
dilaton $\vh$.
The next step is the identification of the conformal quantities
of the Hoyle-Narlikar cosmology (geometric time $T$, distance $D$,
density of the matter-energy $\rho(\vh)=H_0/V_0$, Hubble's parameter
${\cal H}_0(T_0)=\vh'(T_0)/\vh(T_0)$) with the observational ones.

The evolution of the dilaton is determined by equation~(\ref{d56})
\be
\label{d56c}
\vh'\equiv \frac{d\vh}{dT}=\sqrt{\rho(\vh)}~,
\ee
where $\rho(\vh)$ is considered as a measurable density of matter in
the conformal cosmology.
We can express the present-day value of the dilaton $\vh(T_0)$ in terms of
observational quantities: the density $\rho$ and the Hubble parameter
\be
\label{d12a}
\vh(T_0)=\frac{\sqrt{\rho(T_0)}}{{\cal H}_{0}}.
\ee
The cosmological observational data for the density parameter $\Omega
= \rho/\rho_c$ with $\rho_{c}=3{\cal H}_0^2 M_{\rm Planck}^2/(8 \pi)$ 
allow us to assert that the present value of the dilaton field is related to 
the Planck mass,
\be \label{bc}
\vh(T_0)=M_{\rm Planck}\sqrt{\frac{3}{8\pi}}~,
%\Omega^{1/2}
%\approx M_{\rm Planck}\sqrt{\frac{3}{8\pi}}
\ee
for $\Omega \approx 1$, see below.
Thus, on the fundamental level of CGR, the Planck mass
is not a fundamental constant, but determined by the value
of the dilaton field $\vh(T)$~(\ref{bc}).
%\be \label{bc0}
%\vh(T_0)=M_{\rm Planck}\sqrt{\frac{3}{8\pi}}
%\ee
This is a difference of principle between CGR and Einstein's GR.
One can relate both theories by fixing an interval of
the absolute (conformal - noninvariant) world time in Einstein's theory
\be\label{frws}
dT_f = a(T)dT~~~~~     \left(a(T)=\frac{\vh(T)}{M_{\rm Planck}}
\sqrt{\frac{8\pi}{3}}\equiv \frac{\vh(T)}{\vh(T_0)}\right)~.
\ee
Einstein's theory supposes that an observer measures
this absolute interval in Riemannian space. As a result, he
obtains the Friedmann-Robertson-Walker (FRW) cosmology where the
redshift is treated as expansion of the universe and the measurable
density $\rho^{\rm exp}$ is identified with the Einstein
density $\rho_{\rm Einstein}=\rho^{\rm exp}/a^4$.
In the FRW cosmology the experimental fit~(\ref{d12a}) is treated
as a standard definition of the critical density provided $a(T_0)=1$.
The coincidence of the values of the scale
factors in both the versions of cosmology does not mean the
equivalence of their dynamic evolution in  corresponding times.
In the FRW version, the mass density is
decreasing, whereas in the conformal version, the mass density is increasing.

Thus, in CGR, a large wave-length of a photon 
$\lambda_S=[y\vh(T_0-D/c)]^{-1}$ emitted from an atom in a star at
distance $D$ is compared to a small one $\lambda_E=[y\vh(T_0)]^{-1}$
corresponding to an atom on earth within a stationary universe.
In GR, the wave-length of a photon from a distant star
$\lambda_S$ becomes greater by a factor of $a(T_0)/a(T_0-D/c)$
due to the cosmic evolution of all length during its travel to the
observer on Earth.
In both the cases we obtain a redshift $Z>1$.
In CGR, $Z>0$ is explained by the increasing atomic masses.
In GR, $Z>0$ is explained by increasing the star photon wave-length
during its flight.

%\subsection*{The problem of dark matter}

To discuss the problem of dark matter in the conformal cosmology,
we should also take into account
that the present-day observations
reflect the matter density $\Omega(T_0-D/c)$ at the time when the
light was emitted from the cosmic objects.
The mass density was less than at the present-day value $\Omega(T_0)=\Omega_0$
due to the mass increase of the particles by the dilaton field $\vh(T)$ 
with progressing conformal time.
This effect of the retardation in the matter density can be roughly estimated 
by averaging
$\Omega(T_0-D/c)$ over distances (or time) introducing the coefficient
$$
{\gamma}=\frac{{T_0 \Omega_0}}{\int\limits_0^{T_0} dT\Omega(T)}.
$$
For the dust stage the coefficient of the mass increase is $\gamma=3$.
We get for the present-day
value of the cosmic matter density in CGR and the Planck "constant" in GR
\be\label{d32}
\Omega(T_0) = \gamma \Omega_0^{\rm exp} \approx 1,
\ee
where the value of $\Omega_0^{\rm exp} \approx 0.3$ has been taken from 
a recent analysis of the total luminous and dark matter density \cite{turner}.
This result would entail that in a flat universe case, there is no
reason for dark energy, since the missing energy density problem
occurs only due to the neglect of this mass retardation effect in the
standard analysis of cosmological parameters. The final question
arises: Can we account within the conformal cosmology scenario for the 
observed cosmic acceleration without the cosmological constant or
quintessence models?

%\subsection*{The present-day stage: quantum field cosmology}

At the present-day stage, the Bogoliubov quasiparticles
coincide with the measurable particles, so that the measurable density of
energy of matter in the universe is a sum of relativistic energies of all
particles in it (with the number of particles $N_{n,f}(k)$)
\be\label{qfc}
\rho(\vh)=\sum\limits_{f,k_n }^{ }\sqrt{k_n^2+y_f^2\vh^2}
\frac{N_{n,f}(k_n)}{V_0}
\ee
The wave-function of the Universe $\Psi_{\rm univ.}$
is nothing but the
product of oscillator wave-functions
$\Psi_{\rm univ.}=\prod \Psi_{\rm part.}=\prod
\exp({\imath T \sqrt{k_n^2+y_f^2\vh^2}})$.

Neglecting masses ($y_f=0$),
we get the conformal version of the radiation stage ($\rho(\vh)=\bar \rho_R$)
for an observer with the relative standard.
The evolution law for a scalar field in this case is
\be
\varphi(T_{0})=\sqrt{\bar \rho_R}T_{0},
\ee
and the Hubble parameter ${\cal H}_0(T_{0})$ is
\be
{\cal H}_0=\frac{1}{\vh}\frac{d\vh}{d T_{0}}=\frac{1}{T_{0}} .
\ee

Neglecting momenta ($k_n=0$), we get the conformal version of the
dust stage ($\rho(\vh)=\vh\bar \rho_D$) with the evolution law

\be
\varphi(T_{0})=\frac{T_{0}^2}{4}\bar \rho_D~,
\ee
the Hubble parameter ${\cal H}_0(T_{0})$
\be
{\cal H}_0 = \frac{\vh\prime}{\vh}=\frac{2}{T_{0}}~,
\ee
and the acceleration-parameter
\be
q = -\frac{\vh'' \vh}{\vh'^2} = - \frac{1}{2}~,
\ee
which is in agreement with the recent data from the  Supernova
Cosmology Project~\cite{snov}.

\vspace{0.5cm}

\section*{Summary}

In this paper, we have emphasized the conformal-invariant treatment
of the GR dynamics is compatible with the Weyl
geometry of similarity but not with the Riemannian one. The geometry
of similarity converts the conformal-invariant Lichnerowicz
variables from an effective mathematical tool to physical
observables.

The important consequence of the geometry of similarity
is the conformal cosmology.

The present-day value of the dilaton expressed in terms of
observational data of the field version of the Hoyle-Narlikar
cosmology coincides with the Planck mass within
the limits of observational errors.

The conformal cosmology and its quantum version allow us to give
answers to a set of problems of the standard cosmology including
the positive arrow of the geometric time (as a consequence
of the positive energy of a dynamic system and of its stability),
the anisotropic inflation stage,
the creation of a universe as a dynamical system in the field world space,
and as well as the retardation origin of
dark matter (as we estimate the present-day mass density using data
of the earlier stages where the mass density was smaller).
We would like to emphasize that the phenomenon of the accelerating
evolution of the universe in the dust stage is in agreement with the 
present observational data.

\section*{Acknowledgments}

The authors are grateful to
%\medskip
B.~Barbashov,
%A.~Borowiec, A.~Efremov, V.~Papoyan,  
and M.~Pawlowski
%, V.~Smirichinski
for fruitful discussions.
D.B. and D.B. acknowledge support from the Heisenberg-Landau program for 
visiting the JINR Dubna where part of this work has been done.


\begin{thebibliography}{}
\bibitem{pr}
Pawlowski M., and Raczka R., {\it Found. of Phys.} {\bf 24}, 1305 (1994).
\bibitem{grg}
Gyngazov L.N., Pawlowski M., Pervushin V.N., and
Smirichinski V.I., {\it Gen. Rel. and Grav.} {\bf 30}, 1749 (1998).
\bibitem{plb}
Pawlowski M.,  Papoyan V.V., Pervushin V.N., and Smirichinski V.I., 
{\it Phys. Lett. } {\bf B  444}, 293 (1998).
\bibitem{ps1}
Pervushin V.N., and Smirichinski V.I., {\it J. Phys. A: Math. Gen.}
{\bf 32}, 6191 (1999).
\bibitem{6116}
Pawlowski M., and Pervushin V.N.,
{\it Reparametrization-invariant path integral in GR and
"Big Bang" of Quantum Universe}
{Preprint JINR} E2-2000-67 (Dubna);~[hep-th/0006116] (2000). 
\bibitem{kl} Kallosh R., Kofman L., Linde A., and Van Proeyen A.,
{\it Superconformal Symmetry, Supergravity and Cosmology},
Preprint CERN-TH/2000-118, [hep-th/0006179] (2000).
\bibitem{pct} Penrose R., {\it Relativity, Groups and Topology},
Gordon and Breach, London (1964);\\
Chernikov N., and Tagirov E., {\it Ann. Inst. Henri Poincar\`e} {\bf 9}, 109 
(1968).
\bibitem{jbd}
Jordan P., {\it Schwerkraft und Weltall}, Vieweg und Sohn, Braunschweig (1955);
\\
Brans C., Dicke R.H., {\it Phys. Rev.} {\bf 124}, 925 (1961).
\bibitem{we} Weyl H., {\it Sitzungsber. d. Berl. Akad.}, p. 465 (1918).
\bibitem{N}
Narlikar J.V., {\it Space Sci. Rev.} {\bf 50}, 523 (1989).
\bibitem{L}
Lichnerowicz A., {\it Journ. Math. Pures and Appl.} {\bf B 37}, 23 (1944);\\
York J.W.(Jr.), {\it Phys. Rev. Lett.} {\bf 26}, 1658 (1971).
%
\bibitem{bww}
Bogoliubov N.N., and Shirkov D.V., {\it Introduction to the Theory of 
Quantized Fields}, Interscience Publishers, New York (1959).
\bibitem{lp}
Parker G.L., {\it Phys. Rev.} {\bf 183}, 1057 (1969); 
{\it Phys. Rev.} {\bf D3}, 346 (1971).
\bibitem{z} Zel'dovich Ya.B., and Starobinski A.A., {\it Zh.
    Exp. Teor. Fiz.} {\bf 61}, 2161 
(1971).
\bibitem{rt}
Regge T., and Teitelboim C., {\it Ann. Phys.} {\bf 88}, 286 (1974).
\bibitem{ws}
Weinberg S.,{\it Rev. Mod. Phys.} {\bf 61}, 1 (1989).
\bibitem{ppv}
Pervushin V., Proskurin D., and Vinitsky S.,
{\it Conformal-invariant theory of the early universe},
Proc. of Int. Conf. {\it Group Theoretical Methods in Physics}, Dubna,  JINR,
July 31- August 5 (2000).
%\bibitem{rpp}
%{\it Review of Particle Physics}, {\it Phys. Rev.} {\bf D 54}, 107  (1996).
\bibitem{M}
Misner C.,  {\it Phys. Rev.} {\bf 186}, 1319 (1969).
\bibitem{turner}
Turner M.S., {\it Phys. Scripta} {\bf T 85}, 210 (2000). 
\bibitem{snov}
Perlmutter S., et al., {\it Astrophys. J.}
 {\bf 517}, 565 (1999).
%\end{multicols}
\end{thebibliography}
\end{document}